\documentclass[conference]{IEEEtran}
\IEEEoverridecommandlockouts

\usepackage{lineno}
\modulolinenumbers[5]

\usepackage{soul}
\usepackage{amssymb}
\usepackage{amsmath}
\usepackage[utf8]{inputenc}
\usepackage{graphicx}
\usepackage{epstopdf}
\usepackage{nth}
\usepackage{caption}
\usepackage{float}
\usepackage{multicol}
\usepackage[printonlyused,withpage]{acronym} 
\usepackage[disable]{todonotes} 
\usepackage{mathtools}
\usepackage{color}
\usepackage{enumitem}
\usepackage{footnote}
\usepackage{url}
\usepackage{cleveref}
\usepackage{subcaption}
\usepackage[normalem]{ulem}
\IEEEtriggeratref{2}
\usepackage{tikz}
\usepackage{pgfplots}
\usepackage{varwidth}
\usetikzlibrary{plotmarks}
\usetikzlibrary{backgrounds}
\usetikzlibrary{fit}
\usetikzlibrary{graphs}
\usepgfplotslibrary{groupplots}
\usepgfplotslibrary{colorbrewer}	
\usepgfplotslibrary[statistics]
\usepgfplotslibrary{external}

\usetikzlibrary{chains,shapes.multipart}
\usetikzlibrary{shapes,calc,fit,arrows}
\usetikzlibrary{automata, positioning}

\usetikzlibrary{decorations.pathreplacing,decorations.markings,shapes.geometric,arrows,shapes,snakes,automata,backgrounds,petri}

\usetikzlibrary{graphs}

\makeatletter
\def\algbackskip{\hskip-\ALG@thistlm}
\makeatother

\makesavenoteenv{tabular}
\makesavenoteenv{table}

\newcommand{\william}[1]{\todo[inline,size=\small,color=red!30,caption={}]{\textbf{William:} #1}}
\newcommand{\per}[1]{\todo[inline,size=\small,color=blue!30,caption={}]{\textbf{Per:} #1}}

\newacro{5G}{Fifth Generation Wireless Specifications}
\newacro{AP}{Application Providers}
\newacro{API}{Application Program Interface}
\newacro{AR}{Augmented Reality}
\newacro{ARQ}{Automatic Repeat Query}
\newacro{AWS}{Amazon Web Services}
\newacro{BER}{Benefit Effort Ratio}
\newacro{BER}{Bit Error Rate}
\newacro{CAPEX}{Capital Expenditure}
\newacro{CDN}{Content Delivery Network}
\newacro{CLI}{Command Line Interface}
\newacro{CPS}{Cyber-Physical System}
\newacro{CPU}{Central Processing Unit}
\newacro{CRC}{Cyclic Redundancy Check}
\newacro{CSI}{Channel State Information}
\newacro{DB}{Database}
\newacro{DC}{Data Center}
\newacro{DoF}{Degrees of Freedom}
\newacro{DOF}{Degrees Of Freedom}
\newacro{EC2}{Elastic Compute Cloud}
\newacro{FaaS}{Function-as-a-Service}
\newacro{FEC}{Forward Error Correction}
\newacro{FPGA}{Field-Programmable Gate Array}
\newacro{GUI}{Graphical User Interface}
\newacro{HARQ}{Hybrid Automatic Repeat Query}
\newacro{HW}{Hardware}
\newacro{IaaS}{Infrastructure as a Service}
\newacro{i.i.d}{Independent and Identically Distributed random variables}
\newacro{IoE}{Internet of Everything}
\newacro{I/O}{Input/Output}
\newacro{IoT}{Internet of Things}
\newacro{IP}{Infrastructure Providers}
\newacro{LDPC}{Low-Density Parity-Check }
\newacro{LTE}{Long Term Evolution}
\newacro{LuMaMi}{Lund Massive MIMO}
\newacro{MAC}{Medium Access Control}
\newacro{MAN}{Metropolitan Area Network}
\newacro{MCN}{Heterogeneous Distributed Computing}
\newacro{MCN}{Mobile Cloud Network}
\newacro{MD}{Mobile Device}
\newacro{MD}{Mobile Devices}
\newacro{MEC}{Mobile Edge cloud}
\newacro{MIMO}{Multiple Input Multiple Output}
\newacro{MIP}{Mixed Integer Programming}
\newacro{mMTC}{massive Machine Type Communication}
\newacro{MN}{Mobile Network}
\newacro{MNO}{Mobile Network Operator}
\newacro{MNO}{Mobile Network Operators}
\newacro{MPC}{Model Predictive Control}
\newacro{MQTT}{Message Queue Telemetry Transport}
\newacro{MR}{Maximum-Ration Combining}
\newacro{MT}{Mobile Terminal}
\newacro{MU-MIMO}{Multi-User MIMO}
\newacro{NFV}{Network Function Virtualisation}
\newacro{NoOps}{No Operations}
\newacro{OFDM}{Orthogonal Frequency-Division Multiplexing}
\newacro{OPEX}{Operational Expenditure}
\newacro{PaaS}{Platform as a Service}
\newacro{PDC}{Proximal Data Centers}
\newacro{PID}{Proportional Integral Derivative}
\newacro{PM}{Physical Machine}
\newacro{QoS}{Quality of Service}
\newacro{QPSK}{Quadrature Phase Shift Keying}
\newacro{RAN}{Radio Access Network}
\newacro{RAT}{Radio Access Technology}
\newacro{RBS}{Radio Base Station}
\newacro{RBS}{Radio Base Stations}
\newacro{RDC}{Remote Data Centers}
\newacro{RTD}{Round-Trip Delay time}
\newacro{RTT}{Round Trip Time}
\newacro{RTT}{Round-Trip Time}
\newacro{SaaS}{Software-as-a-Service}
\newacro{SDK}{Software Development Kit}
\newacro{SDN}{Software Defined Networks}
\newacro{SDR}{Software Defined Radio}
\newacro{SLA}{Service Level Agreement}
\newacro{SLO}{Service Level Objective}
\newacro{SLO}{Service Level Objectives}
\newacro{SNR}{Signal-to-Interference-plus-Noise Ratio}
\newacro{SNS}{Simple Notification Service}
\newacro{SoS}{System of Systems}
\newacro{SP}{Service Providers}
\newacro{SQL}{Structured Query Language}
\newacro{SUMO}{Simulation of Urban MObility}
\newacro{SW}{Software}
\newacro{TLS}{Transport Layer Security}
\newacro{TraCI}{Traffic Control Interface}
\newacro{TSC}{Traffic Signal Control}
\newacro{TSP}{Transit Signal Priority}
\newacro{TTI}{Transmission Time Interval}
\newacro{UE}{User Equipment}
\newacro{UM}{User Mobility}
\newacro{URLLC}{Ultra-Reliable and Low-Latency Communication}
\newacro{UX}{User Experience}
\newacro{WAN}{Wide Area Network}
\newacro{WLAN}{Wireless Local Area Network}
\newacro{VM}{Virtual Machine}
\newacro{WSN}{Wireless Sensor Network}
\newacro{vSoftPLC}{virtual Software Programmable Logic Controllers}
\newacro{ZF}{Zero-Forcing}
\newacro{MS}{Mobile Station}
\newacro{NTP}{Network Time Protocol}
\newacro{PTP}{Precision Time Protocol}
\newacro{NGCC}{Next Generation Cloud Computing}
\newacro{ERDC}{Ericsson Research Data Center}
\newacro{DNR}{Distributed-NodeRED}
\newacro{ADC}{Analog to Digital Converter}
\newacro{DAC}{Digital to Analog Converter}
\newacro{NoOps}{No-Operations}
\newacro{PaaS}{Platform-as-a-Service}
\newacro{WASP}{Wallenberg Autonoms Systems and Software Program}

\newcommand{\paradigm}{edge cloud}

\newcommand{\process}{ball and beam}
\newcommand{\cnode}{compute node}

\tikzset{naming/.style={align=center,font=\small}}
\tikzset{antenna/.style={insert path={-- coordinate (ant#1) ++(0,0.25) -- +(135:0.25) + (0,0) -- +(45:0.25)}}}
\tikzset{station/.style={naming,draw,shape=dart,shape border rotate=90, minimum width=10mm, minimum height=10mm,outer sep=0pt,inner sep=3pt}}
\tikzset{mobile/.style={naming,draw,shape=rectangle,minimum width=12mm,minimum height=6mm, outer sep=0pt,inner sep=3pt}}
\tikzset{radiation/.style={{decorate,decoration={expanding waves,angle=90,segment length=4pt}}}}

\tikzset{
reservoiri/.pic={
  \draw[line width=1pt]
    (0,0.25) --++ (0.5, -0.25) -- ++(0.5,0) -- ++(0,-0.5) -- ++(-0.5,0) -- ++(-0.5,-0.25);
   \node[above] at (0.5, 0.5) [text width=3cm, align=center] {#1};
   \coordinate (-input) at (0,-0.25);  
   \coordinate (-output) at (1,-0.25);  
  },
queuei/.pic={
  \draw[line width=1pt]
    (0,0) -- ++(1,0) -- ++(0,-0.5) -- ++(-1,0);
   \foreach \Val in {1,...,4}
     \draw ([xshift=-\Val*5pt]1,0) -- ++(0,-0.5);
   \node[above] at (0.5,0) {#1}; 
   \coordinate (-input) at (0,-0.25);  
   \coordinate (-output) at (1,-0.25);  
  },
rbs/.pic={
  \node[station] (base) {};

  \draw[line join=bevel] (base.100) -- (base.80) -- (base.110) -- (base.70) -- (base.north west) -- (base.north east);
  \draw[line join=bevel] (base.100) -- (base.70) (base.110) -- (base.north east);

  \draw[line cap=rect] ([xshift=-.1768cm,yshift=.6pt]base.north -| base.right tail) [antenna=1];
  \draw[line cap=rect] ([yshift=.6pt]ant1 |- base.north) -- node[above,shape=rectangle,inner ysep=+.3333em]{\dots} ([xshift=.1768cm,yshift=.6pt]base.north -| base.left tail) [antenna=2];

  \draw[thick,radiation,decoration={angle=45}] (.5,1.25) -- +(45:0.5);
  \draw[thick,radiation,decoration={angle=45}] (-.5,1.25) -- +(-45:-0.5);

  },
block/.pic = {
  \draw [line width=1pt] (0,0) rectangle (0.5,-0.5) node[pos=.5] {#1};
  \coordinate (-input) at (0,-0.25);  
  \coordinate (-output) at (0.5,-0.25); 
  }
}
\tikzset{naming/.style={align=center,font=\small}}
\tikzset{antenna/.style={insert path={-- coordinate (ant#1) ++(0,0.25) -- +(135:0.25) + (0,0) -- +(45:0.25)}}}
\tikzset{station/.style={naming,draw,shape=dart,shape border rotate=90, minimum width=10mm, minimum height=10mm,outer sep=0pt,inner sep=3pt}}
\tikzset{mobile/.style={naming,draw,shape=rectangle,minimum width=12mm,minimum height=6mm, outer sep=0pt,inner sep=3pt}}
\tikzset{radiation/.style={{decorate,decoration={expanding waves,angle=90,segment length=4pt}}}}

\begin{document}

\title{\huge{Towards Mission-Critical Control at the Edge and Over 5G}
}

\author{\IEEEauthorblockN{Per Skarin\IEEEauthorrefmark{1}\IEEEauthorrefmark{2}\IEEEauthorrefmark{3}, William Tärneberg\IEEEauthorrefmark{1}\IEEEauthorrefmark{4}, Karl-Erik Årzen\IEEEauthorrefmark{2}, and Maria Kihl\IEEEauthorrefmark{4}}
\IEEEauthorblockA{\IEEEauthorrefmark{1}These authors contributed equally to this work}
\IEEEauthorblockA{\IEEEauthorrefmark{2} Department of Automatic Control, Lund University, Sweden}
\IEEEauthorblockA{\IEEEauthorrefmark{3}Ericsson Research, Lund, Sweden}
\IEEEauthorblockA{\IEEEauthorrefmark{4}Department of Electrical and Information Technology, Lund University, Sweden}
}

\maketitle

\begin{abstract}

With the emergence of industrial IoT and cloud computing, and the advent of 5G and edge clouds, there are ambitious expectations on elasticity, economies of scale, and fast time to market for demanding use cases in the next generation of ICT networks.
Responsiveness and reliability of wireless communication links and services in the cloud are set to improve significantly as the concept of edge clouds is becoming more prevalent.
To enable industrial uptake we must provide cloud capacity in the networks but also a sufficient level of simplicity and self-sustainability in the software platforms. 
In this paper, we present a research test-bed built to study mission-critical control over the distributed edge cloud. We evaluate system properties using a conventional control application in the form of a Model Predictive Controller.
Our cloud platform provides the means to continuously operate our mission-critical application while seamlessly relocating computations across geographically dispersed compute nodes.
Through our use of 5G wireless radio, we allow for mobility and reliably provide compute resources with low latency, at the edge.
The primary contribution of this paper is a state-of-the art, fully operational test-bed showing the potential for merged IoT, 5G, and cloud. We also provide an evaluation of the system while operating a mission-critical application and provide an outlook on a novel research direction.

\end{abstract}


\section{Introduction}
\begin{figure*}[t!]
		\centering
		\includegraphics[width=\linewidth]{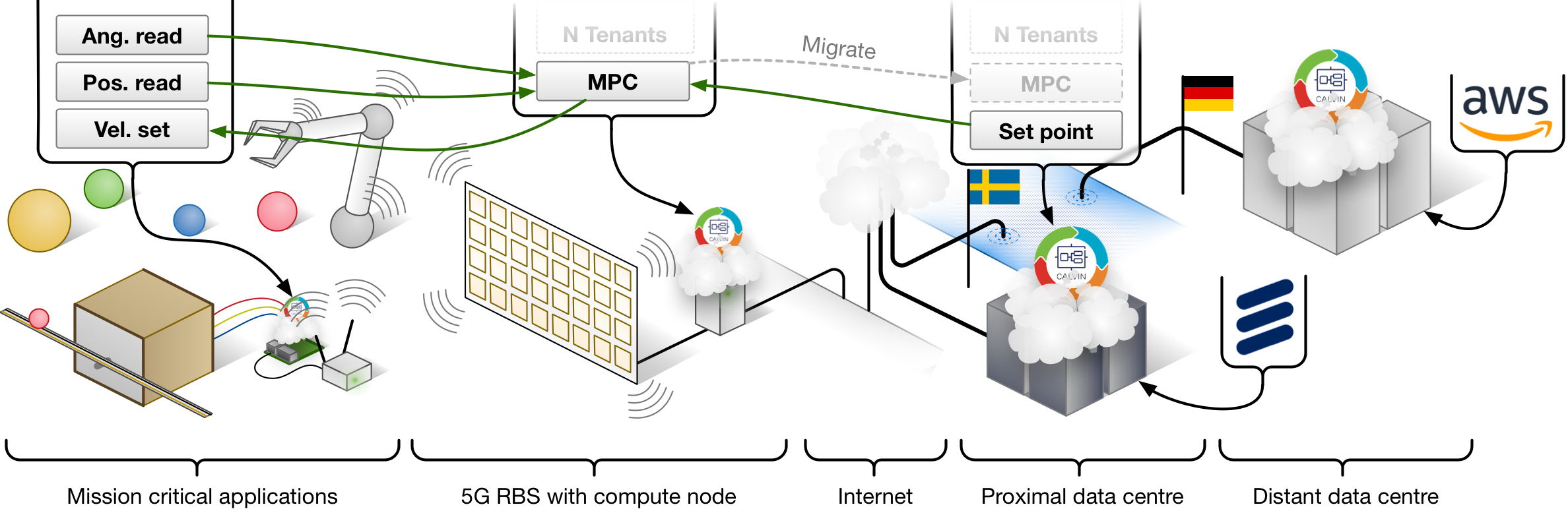}
		\caption{System overview}
		\label{fig:overview}
	\end{figure*}

	In this paper we target the feasibility of running time-sensitive and mission-critical applications in the \paradigm{}. 
	We proceed by designing and building an \paradigm{} research test-bed that encompasses a distributed set of compute nodes, a distributed \ac{PaaS} framework, a \ac{5G} cell, and a time-sensitive mission-critical process under control, see \Cref{fig:overview}.
	
	A mission-critical system is one in which a failure or interruption comes with an unacceptable business or human cost. 
	Here, a failure may be an arbitrarily small deviation from the desired operation.
	Naturally, this includes all systems that create a risk of injury but also systems in which failure incurs a very notable inconvenience, such as a means of transportation rendered practically useless because it is making passengers nauseous.
	Such applications are time-sensitive in the manner that they are unable to cope with delay and jitter in the delay, to the point where they violate their requirements. 
	
	The \acp{WAN} separating a time-sensitive mission-critical system from  a traditional distant \ac{DC} may incur latencies beyond what is operationally acceptable. 
	The \paradigm{} was proposed to mitigate the latent latency, throughput, and availability barriers that separate the end users from distant \acp{DC}.
	When accessing the \paradigm{} over \ac{URLLC} \ac{5G} \cite{shafi20175g}, the latencies are sufficiently low that time-sensitive mission-critical applications can be deployed in the \paradigm{}.
	An additional benefit of the \paradigm{} is that resident applications can be made spatially redundant and fall-back solutions can be implemented at various geographical points in the infrastructure for additional resilience.
	
	Historically, control systems have been deployed as monolithic \ac{SW} implementations on carefully tuned \ac{HW}, adjacent to the plants they control.
	Deploying monolithic \ac{SW} on static \ac{HW} makes such systems undesirably non-modular, less extensible, and limits their ability to self-adapt.
	Conversely, cloud-native applications are built for the cloud, offering the prospect of greater flexibility, reuse, availability, and reliability with lower latencies.
	When applications are implemented in a disaggregated manner, their execution can be distributed across the system's many nodes, migrated, and scaled to meet their individual objectives as well as that of the system as a whole.
	To adapt to, and prosper in, the \paradigm{}, applications will arguably have to adhere to a cloud-native paradigm.
	
	The premise of this paper is that deploying mission-critical applications over the cloud, with wireless devices, must arguably occur in conjunction with the availability of edge cloud resources, the flexibility of cloud-native applications, and the reliability and low latency of \ac{5G}.
	We argue that such applications can operate in and make use of a distributed \paradigm{} but that there need to be relevant tools for them to be native to this context.
	There are many challenges and performance uncertainties in this premise. 
	Therefore, we study the feasibility of deploying time-sensitive mission-critical applications and their performance when deployed on an actual \paradigm{} infrastructure.
	The contributions of this paper are:
	
	\begin{itemize}
		\item A state-of-the art \paradigm{} research test-bed aimed at the study of software autonomy and mission-critical applications.
		\item An empirical baseline evaluation of the plausibility of deploying latency-sensitive applications in the \paradigm{}.
		\item An empirical evaluation of the system's ability to dynamically reconfigure during run-time and the impact this has on the application.
		\item An empirical evaluation of the benefits of deploying latency-sensitive mission-critical applications at the edge, at the plant, and on a distant \ac{DC}.
	\end{itemize}

	\Cref{sec:related_works} covers the related work in the field and highlights the research gap.
	This is followed by a detailed account of the implemented test-bed in \Cref{sec:implementation}. 
	\Cref{sec:evaluation} presents an example automatic control application which is used to evaluate the test-bed.
	Finally, \Cref{sec:conclusions} highlights the contributions of the paper and points to new research directions.
\section{Related work} \label{sec:related_works}
	In this section, we survey related work and highlight the apparent research gaps left by the literature.
	The works below cover both related attempts at research test-beds and experiments that pursue the viability of the \paradigm{}.
	
	Test-beds spanning \acp{UE}, wired and wireless networks, distributed cloud infrastructure and platforms are crucial instruments to realise and study the complexity of the \paradigm{}. 
	The literature contains a number of such attempts.
	The authors of \cite{kang2013savi} present the SAVI test-bed which is an edge cloud test-bed realising \ac{NFV} with a \ac{FPGA}-cloud. 
	SAVI is used in the investigation of virtualising the wireless access network. 
	Although comprehensive, the test-bed does not provide a general edge cloud implementation for cloud native applications nor does the implementation span multiple tiers of cloud resources, including the device. 
	In \cite{7467436}, a full test-bed using existing wireless technologies, \ac{IoT} frameworks, and devices is deployed on an actual production line.
	The industrial applications targeted in the paper are not time-sensitive and the focus is on framework integration rather than system and application performance.
	The authors of \cite{Hu:2016:QIE:2967360.2967369} implemented a rudimentary edge cloud test-bed to quantify the impact of Edge Computing on Mobile Applications using WiFi and the public 4G network. 
	Their effort reveals significant latency and energy usage improvements compared to distant \acp{DC}.
	Additionally, we have in a previous work \cite{Tarneberg:2016:ECF:2996890.2996911} studied the performance of cloud native applications on commercially available platforms in a smart city context.
	
	iFogSim \cite{gupta2017ifogsim} offers a platform for abstract modeling of resource management techniques in \ac{IoT} and edge cloud environments, through simulation.
	In previous works we have also developed several simulators for studying the dynamics of the edge cloud, culminating in \cite{tarneberg2017distributed}. 
	In the absence of a test-bed and in the pursuit of greater flexibility at a higher level of abstraction, simulators are valuable tools.
	However, we are at the point where a test-bed can be practically implemented and where a simulator cannot capture the complexities of an \paradigm{}.

	Other works attempt to characterise and profile the performance of different aspects of the \paradigm{}.
	For example, the authors of \cite{mahmud2014evaluating} evaluate a generic platform for industrial control, with respect to latency, throughput, and CPU load. 
	Their focus is on the pros and cons of virtualisation in a 'Multi-core' environment rather than the 'cloud'. 
	Similarly, \cite{horn2016feasibility} implements a water tank control process and evaluates latency over a \ac{vSoftPLC} on top of LinuxRT.
	In both of these works, the system implementation rather than the plant under control is evaluated. 
	In \cite{hegazy2015industrial}, the authors propose \textit{Industrial automation as a cloud service}. 
	In the paper the authors evaluate a system of time-sensitive control processes in a 1-tier distributed cloud environment. 
	Latency compensation is modelled and redundancy with stability and smooth controller handover is achieved. 
	True for all systems surveyed above, is that they are neither mission-critical nor time-sensitive at the scale addressed in this paper. 

\section{Research test-bed} \label{sec:implementation}	
	Our research test-bed consists of a 5G radio \textit{transmitter} and multiple \textit{receivers} (\acp{UE}) (constituting a \ac{5G} cell), PC-type \textit{\cnode{}s} in \acp{DC} and adjacent to the radio transmitter (the edge node), and reduced capacity \textit{input-output} devices and \textit{physical plants} at the radio receiver ends.
	Here, the physical plant can be a mechanical device that continuously performs a task, for example a robotic arm or an autonomous vehicle.
	We assume that the process which controls the plant is mission-critical and time-sensitive. 
	
	We desire a platform with enough knowledge about the application to perform load balancing while allowing an application its own mobility within the network of compute nodes.
	Further, there is a strong interplay between the edge cloud and the end user equipment such that an application can automatically scale on top of the cloud and provide fall back on local devices.
	The radio subsystem shall be capable of parallel, synchronized low latency communication with several receivers.
	We proceed by defining the properties of the system's individual components.
	From here on the research test-bed is referred to as \textit{the system}.

	\subsection{5G}
		A \ac{5G} wireless system represents the next generation wireless infrastructure \cite{shafi20175g}.
		The emerging focus of \ac{5G} is \ac{URLLC} and \ac{mMTC} where a large number of \ac{IoT} devices, can reliably be served simultaneously at a low latency, $\leq 5 ms$.
		These conditions cannot be replicated with current 802.11 or \ac{LTE} systems. 
		
		A next generation wireless network also implies a deeper integration with associated cloud computing resources. 
		On-demand resources are integrated into the \ac{RBS} and access networks to off-load the back-haul and eliminate the latency overhead of traversing multiple networks and providers.
		
		Massive \ac{MIMO} is the emerging \ac{RAT} for \ac{5G}.
		Fundamentally, massive \ac{MIMO} is a \ac{MU-MIMO} scheme, which can simultaneously communicate with multiple \acp{UE} on the same wireless resource. 
		Additionally, on the \ac{RBS}-side, massive \ac{MIMO} operates with significantly more antennas than existing \ac{LTE}-based \acp{RAT}.
		Massive \ac{MIMO} is typically configured with an order of magnitude more \ac{RBS}-side antennas than simultaneously served \acp{UE}.
		Consequently, the system's spectral efficiency is a few orders of magnitude greater than existing \acp{RAT}.
		The increased spectral efficiency can be used towards serving more simultaneous \acp{UE}, increase throughput, or realising \ac{mMTC}, beyond what can be achieved with existing \acp{RAT}.\per{Eventuellt skulle här till något om 'reliability'}
		
		We implement a \ac{5G} wireless network using \ac{LuMaMi}.
		\ac{LuMaMi} is a Massive \ac{MIMO} test-bed at Lund University, Sweden.
		\ac{LuMaMi}'s scope and detailed implementation are found in \cite{malkowsky2017world}.
		\ac{LuMaMi} can be seen as one solitary \ac{5G} cell that can simultaneously communicate with twelve \acp{UE}.
		
		\ac{LuMaMi} is configured according to \cite{tarneberg2017utilizing}.
		This configuration premiers low latency and reliability over high throughput.
		The modulation scheme is QPSK.
		The resulting throughput is 4.6 Mbps downlink and 9.1 Mbps uplink, per \ac{UE}, which is more than sufficient to support our application. 
		\ac{LuMaMi} allows us to directly route traffic through the system, allowing us to place a \cnode{}  in the \ac{RBS}.

	\subsection{Edge cloud and network}\label{sec:syssetup}	
		\begin{table}[t]
			\centering
			
			\begin{tabular}{|p{0.1\columnwidth}|p{0.3\columnwidth}|p{0.27\columnwidth}|}
				\bf{Node} & \bf{Device} & \bf{Location} \\ \hline
				Plant &  Raspberry Pi 3Bs & Plant adjacent\\ \hline 
				Edge &  Intel Core i7 Desktop & LuMaMi adjacent \\ \hline
				ERDC & Intel Core i7 VM &  Lund, Sweden. \\ \hline
				AWS &  Intel Xeon VM & Frankfurt, Germany \\ \hline
			\end{tabular}
		
			\caption{Node types}
			\label{tbl:nodes}
		\end{table}

		The system as a whole is tied together by a set of \cnode{}s joined by a network.
		A summary of the \cnode{}s is presented in \Cref{tbl:nodes}.
		The system's network is conceptually configured as depicted in \Cref{fig:overview}.
		Adjacent to each plant is a Raspberry Pi. 
		In order to sample and manipulate the plant, each Raspberry Pi has been equipped with a ADC/DAC shield. 
		They are \cnode{}s and may service other functions in addition to interacting with the plant.
		Each Raspberry Pi is also connected to a \ac{5G} \ac{UE}.
		The \ac{5G} cell is isolated in its own subnet.
		The subnet includes the wireless infrastructure, an edge cloud node, and plant nodes.
		The plant-adjacent Raspberry Pis are connected to the system's subnet over \ac{LuMaMi}.
		The wireless edge cloud node is adjacent to the \ac{LuMaMi} \ac{RBS}.
		It therefore connects directly to the \ac{RBS} without traversing additional networks.
		
		A router is connected to the cell's subnet and the larger \acp{DC}.
		The \acf{ERDC} resides in Lund, Sweden a few kilometres from the cell.
		\ac{ERDC} is a research \ac{DC} operated by Ericsson (Lund, Sweden), that is open to industrial and academic research efforts within the \ac{WASP}.
		We run on top of Open Stack Pike and our instance (a c4m16) has four Intel i7 cores registered by Linux as 1.6 Ghz, and 16 GB of RAM.    
		The \ac{AWS} EC2 instance (a c4.large) is hosted on eu-central-1 (Frankfurt, Germany).
		This node has two Intel Xeon cores at 2.9 Ghz and 8 GB of RAM. 
		We do not use all cores and therefore expect the latter system to be the best performing.
		The two \acp{VM} on \ac{ERDC} and \ac{AWS} connect to the subnet over VPN, allowing direct access between all \cnode{}s.

	\subsection{Cloud native application framework}
		We define a cloud-native application to be an application that has been disaggregated into logical and independent components connected in a \textit{data-flow graph} and that is hosted on a \ac{PaaS} framework.
		The \ac{PaaS} and its resident applications ubiquitously operate over multiple geographically distributed and heterogeneous \cnode{}s.
		An application's data flow graph can be rerouted and extended in run-time, when for example adding a new feature.
		Additionally, the components shall be able to traverse the cloud and associate with and discover physical input-output devices if the application so requires.

		Amazon's AWS, Microsoft's Azure, IBM's Bluemix, and Google's Cloud offer their own flavours of cloud native application platforms, ranging from \ac{SaaS} to server-less \ac{FaaS}. 
		However, none of these providers allow their users to define logical data flows nor do they provide necessary performance guarantees.
		They are typically intended for lifting data from the edge and \ac{IoT}-devices to the cloud.
		The services do not provide native support for closing logical loops from edge devices over the cloud and back to the edge device with guarantees on latency and consistency.
		For the purpose of building an open research test-bed, these platforms are proprietary and cannot be arbitrarily deployed and independently managed across an \paradigm{} infrastructure.
		This is a requirement for our system in order to realize the view where one specification of the software can be deployed anywhere.

		In this work, we use Calvin \cite{persson2015calvin} as our cloud platform.
		Calvin is distributed, event-driven, server-less, and is based on a data-flow programming model.
		There are a number of such platforms for different workloads, such as: Nebula \cite{ryden2014nebula}, Node-RED\cite{nodered}, IEC 61499 \cite{vyatkin2011iec}, and Naiad \cite{murray2013naiad}.
		The aforementioned systems are targeted for the \ac{IoT} domain and cater for workloads varying from simple event-driven automation to high-throughput Hadoop jobs, but none of them have been built with the intention to run tight control loops over a dynamic distributed system.
		 
		Of the above, Calvin is most similar to Node-RED. 
		However, while Node-RED emphasizes the programming model and graphical tools, Calvin puts more focus on runtime dynamics and distributed deployment.
		The perspective of Calvin is well attuned to the presentation of the Distributed Data-flow model in \cite{7356560}, where a \ac{DNR} extension is proposed.
		A notable operational difference is that \ac{DNR} employs duplication to realize mobility while Calvin's code migration technique is arguably more efficient and is better suited for computationally intense applications \cite{7356560}.
		Additionally, in Calvin, we can migrate using various optimisation criteria to provide for instance load balancing or jitter reduction.

		Calvin is conceptually structured as follows.
		The operational units of Calvin are called \textit{actors} (nodes in data-flow) while a \textit{runtime} is an instantiation of the Calvin application environment on a device.
		In our present implementation there is a one-to-one mapping between Calvin runtimes and compute nodes and we therefore interchangeably refer to them simply as nodes.
		An actors' input and output messages, are known as \textit{tokens}.
		A set of actors and their interconnections constitute an \textit{application}. 
		Each node independently schedules its resident actors in a round-robin manner.

		Actors' states can be migrated and horizontally scaled across nodes.
		What constitutes an actors' state is defined by the developer. 
		The Calvin framework can autonomously migrate and place actors to load-balance nodes and to meet its own performance goals.
		However, application owners can specify requirements for actors which tie them to a preferred runtime.
		For example, a sensor reading actor can be required to be placed on the node associated with the physical plant it is observing.

\section{Evaluation}\label{sec:evaluation}
	In this section, we present an automatic control application as our time-sensitive mission-critical application.
	The controller is used to evaluate the performance and plausibility of the test-bed.
	That is, can we do mission-critical control over the \paradigm~ and does the system exhibit the properties associated with an \paradigm{}.
	We begin by detailing the application in terms of the software implemented on top of Calvin and the plant that it controls.
	We then evaluate its performance in a set of experiments designed to:
	
	\begin{enumerate}
		\item Reveal the characteristics of the system and the controller by establishing a baseline observation of the performance and behaviour of the controller over long time periods.
		\item Verify the adaptability of the system by continuously migrating the controller actor across the system's nodes, in run-time. 
		\item Explore operating limits of the system's nodes and thus their relative advantage by deploying a well-tuned but computationally demanding and time-sensitive controller on the system.
	\end{enumerate}

	In order to observe the system's performance potential, we choose to limit the study to normal operating conditions.
	Notably, the connections to the \acp{DC} may at times degrade.
	We assume these to be infrequent, transient behaviours and do not consider how to handle them in this work.

	\subsection{Control application} \label{subsec:process}
		We employ a \process{} process \cite{virseda2004modeling} as our plant under control.
		The control has to be fast and there are clear limits set by physical constraints, yet enough flexibility for us to modify conditions to create various operating scenarios. \william{Various -> corner cases? taxing scenarios? \dots}
		Control is critical in that a failure may cause an unrecoverable state.
		
		The objective of the \process{} process is to expediently move to and maintain a ball on a set location (the set-point) on a beam.
		The length of the beam is 110 cm.
		The controller acts on the beam which is manipulated by a motor.
		The plant \textit{outputs} the angle ($\alpha$) of the beam and the position of the ball on the beam ($x$).
		The control signal, the \textit{input} to the plant, is the radial velocity of the beam ($\omega$).
		Naturally, the further we go towards the end of the beam the higher the risk that the ball falls off the beam due to network delays, noisy sensors readings, and other system deficiencies.
		
		
		\tikzexternalenable
		\begin{figure}[t!]
			\centering
			\usetikzlibrary{chains,shapes.multipart}
\usetikzlibrary{shapes,calc,fit,arrows}
\usetikzlibrary{automata, positioning}
\usetikzlibrary{arrows.meta}

\usetikzlibrary{decorations.pathreplacing,decorations.markings,shapes.geometric,arrows,shapes,snakes,automata,backgrounds,petri}

\usetikzlibrary{graphs}

\begin{tikzpicture}[node distance=1.75cm,>=stealth',bend angle=45,auto]

	\filldraw[fill=gray!13!white, draw=white] (1.8,-0.8) -- (7.6,-0.8) -- (7.6,3.6) -- (5.5,3.6) -- (5.5,1) -- (1.8,1) -- cycle;

    \tikzstyle{actor}=[rectangle,rounded corners,thin,draw=black,fill=blue!20,minimum size=6mm, minimum width=1.5cm]
    \tikzstyle{plant}=[rectangle,draw=red!75,fill=red!30]
    \tikzstyle{transition}=[rectangle,thick,draw=black!75,fill=black!20,minimum size=4mm]
    \tikzstyle{arrow}=[->,thick,shorten <=2pt, shorten >=2pt]
    \tikzstyle{port}=[>=Latex, minimum height=1cm]

    \tikzstyle{every label}=[red]

    \node [plant, minimum size=1.5cm] (plant) {Plant};
    \node [actor, fill=green!10] (clk)  [right =of plant] {Clock};
    \node [actor, fill=orange!25] (pos)  [above =of clk] {ADC};
    \node [actor, fill=orange!25] (ang)  [below =of clk] {ADC};
    \node [actor, fill=green!10] (mpc)  [right =of clk] {MPC};
    \node [actor, fill=green!10] (ref)  [above =of mpc] {Set-point};
    \node [actor, fill=orange!25] (vel)  [below =of mpc] {DAC};

    \path[->,every loop/.style={looseness=4}] (ref) edge  [blue,in=70,out=90,loop] node {$N s$} (); 
    \path[->,every loop/.style={looseness=4}] (clk) edge  [blue,in=120,out=100,loop] node [left] {$50ms$} (); 

    \draw [port,->] (clk) edge [thick] node {\textit{tick}} (pos);
    \draw [port,->] (clk) edge [thick] node {\textit{tick}} (ang);

    \draw [port,->] (ref) edge [thick] node {$y_{ref}$} (mpc);

    \draw [->] (plant) edge node {pos} (pos);
    \draw [->] (plant) edge node {ang} (ang);

    \draw [port,->] (pos) edge [thick] node {$y$} (mpc);
    \draw [port,->] (ang) edge [thick] node [below] {ang} (mpc);

    \draw [port,->] (mpc) edge [thick] node {$u$} (vel);
    
    \draw [->] (vel) edge [bend left, out=60] node {$\omega$} (plant);

\end{tikzpicture}
			\caption{Calvin MPC implementation}
			\label{fig:ctrl_impl_mpc}
		\end{figure}
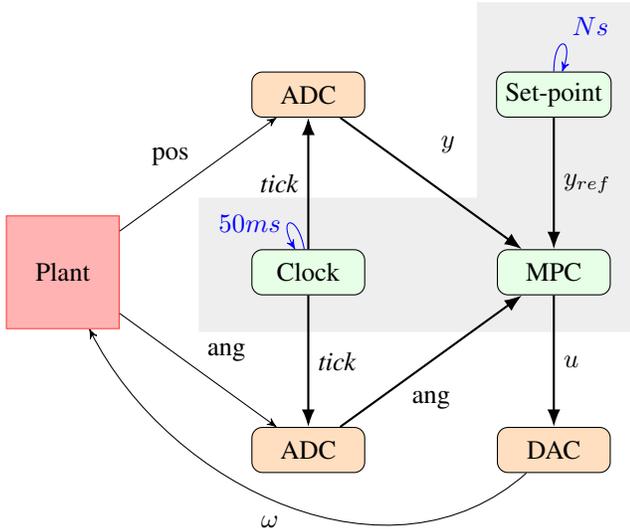
		\tikzexternaldisable
			
		We implement a controller for the \process{} process using \ac{MPC}~\cite{rawlings2010model}. 
		The application periodically samples the position of the ball and the angle of the beam.
		With every sample the \ac{MPC} interacts with the plant by changing the velocity of the beam.
		To figure out what velocity to set the \ac{MPC} performs a numerical optimisation where it takes into account a series of actions that will bring the ball into the desired state.
		In a primitive and general form this optimisation may be expressed as
	
		\begin{align}
			\underset{u_0,u_1,\dots}{minimize} \quad & 	\sum^{T-1}_{t=0} L(x_t, u_t) + \phi(x_T) \label{eq:mpc_min} \\
			\textit{s.t.} \quad & x_{t+1} = f(x_t, u_t) \label{eq:mpc_mod}\\
			& u_t \in U, x_t \in X \label{eq:mpc_constraints}
		\end{align}
		
		where $L(x_t, u_t)$ in \Cref{eq:mpc_min} is the \textit{cost} function which puts a value to a state $x_t$ and control input $u_t$ at time step $t$.
		The function $\phi(x_T)$ assigns a different value specification to the final (or terminal) state $x_T$.
		The number of time steps $T$ is called the horizon and specifies how far into the future the controller anticipates control actions.
		$f(x_t, u_t)$ in \Cref{eq:mpc_mod} is the plant model which specifies the dynamics of how the system states evolve with the time step $t$.
		\Cref{eq:mpc_constraints} is a set of expressions which state limitations to the plant inputs and the state space.
		All of this is defined to set the operating conditions for the controller.
		The initial state, $x_0$, is drawn from measurements and state estimation.
		The end result is a quadratic program which is re-evaluated every sample.
		
		For the optimisation, we use a dynamically linked binary created with the use of QPgen~\cite{qpgen}.
		In our implementation, we use a sampling period of 50 ms (20 Hz).
		This choice is a reasonable trade-off between control performance and early observations of the system's latencies. 
		Unlike the well known and often used \ac{PID} controller, the execution of an \ac{MPC} is demanding and the execution time is not constant.
		The time it takes to solve its optimisation routine varies with disturbances acting on the system and where within its operating range it is currently acting.
		QPgen is an efficient solver and our optimisation problem has few variables, yet we shall see that the time it takes to find a solution can be considerable.
		Sometimes there is no solution or one is very hard to find. \william{Very hard -> needs a significant amount of time?}
		In such a case we end the search after a fixed amount of iterations in the optimiser.
		
		The processor and memory demands of the \ac{MPC} optimization may be considerable and the many ways of tuning it for various situations make it interesting as an \paradigm~application.
		A number of controllers can be designed for the same problem where computational and memory demands are weighted to performance, operational range, stability regions, and erratic behaviour.
		To handle the presence of plant and sensory noise we include a Kalman filter. The filter is also used to estimate the speed of the ball.
		We use a simple plant, a basic \ac{MPC} controller and a standard state estimator but even our rather simple case allows us to study and demonstrate behaviour in the experimental platform and the effects on the control.
		
		The Calvin application graph for the \ac{MPC} control loop is shown in \Cref{fig:ctrl_impl_mpc}. 
		The rounded rectangles represent individual components, implemented as actors, which are deployed onto the systems. 
		The two \ac{ADC} sensory actors adhere to component reuse and the principle idea that they need not be collocated. 
		However, they are to be read jointly and therefore share a clock tick. 
		The components within the grey area can be freely placed within the system. The \acp{ADC} (the position and angle sensors) and the \ac{DAC} (the motor actuator) have an affinity to the plant-adjacent node.
		
		We note that the scheduling and inter-node communication in the software platform introduce a significant amount of delay and jitter, which affects the performance of the controller and cause oscillations.		
		In extension, much can be done in terms of model and controller tuning, state estimation, delay prediction, system improvements etc., but in this work we are interested in studying the overall performance of the platform.
		Remedies and improvements are left for later work, here we characterise the basic system and do not focus on details of control performance.

		All related software runs on top of Linux and the Calvin runtime is launched using real-time priority and the POSIX FIFO scheduling policy.
		We allow the edge node to take full advantage of this as it runs on bare metal.
		The kernels have not been patched with the PREEMPT\_RT patch set \cite{rtlinux}.

	\subsection{System characteristics}\label{ref:character}
		To characterise and verify the basic functionality of the system we ran our \ac{MPC} on each of the nodes in \Cref{sec:syssetup}.
		With each test we let the \ac{MPC} control the beam for 60 minutes while alternating the set-point of the ball between the centre position and one side of the beam.
		To be robust in this experiment, we restrict the set-point with a large margin to the end of the beam.
		On the other hand, the further out we move the ball the more we put the controller to work, which is something we return to in \Cref{sec:tight}.
		
		\tikzexternalenable
		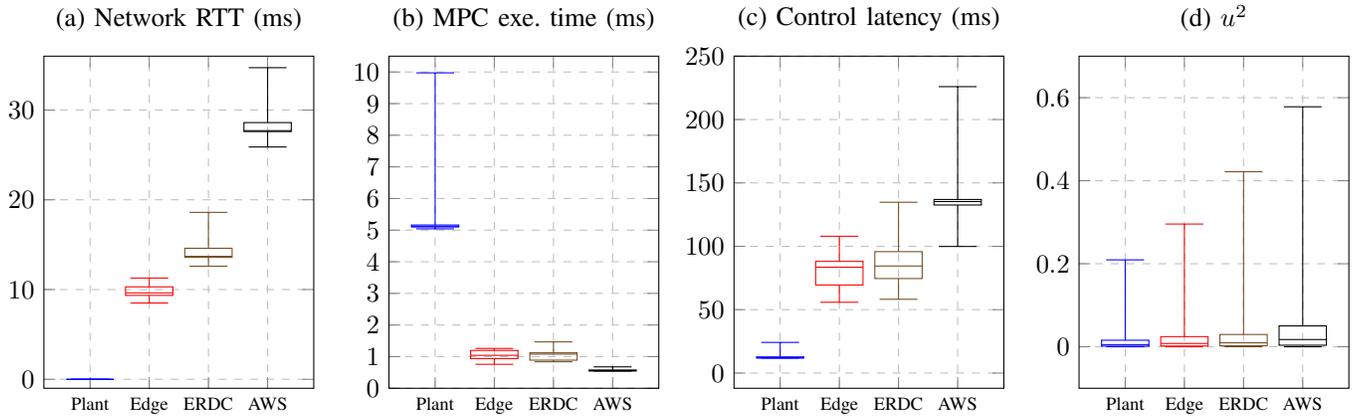
\begin{figure*}[t!]
			\centering 
			\begin{tikzpicture}[transform shape]
    \pgfplotsset{width=0.285\linewidth,height=6.0cm}
    \pgfplotsset{set layers}
    \pgfplotsset{filter discard warning=false}
    \pgfplotsset{grid style={dashed}}

    \begin{groupplot}
    [
    group style={
           group name=my plots,
           group size=4 by 1,
           xlabels at=edge bottom,
           xticklabels at=edge bottom,
           vertical sep=0.3cm,
           },
           ylabel style = {align=center},
    xmajorgrids,
    ymajorgrids,
    xtick={1,2,3,4},
    xticklabels={\scriptsize{Plant}, \scriptsize{Edge}, \scriptsize{\ac{ERDC}}, \scriptsize{\ac{AWS}}},
    scaled ticks=false,
    ]

	%
	%
	\nextgroupplot[
		boxplot/draw direction=y,
		ymajorgrids,
		xmajorgrids,
		ymax=36, ymin=-1,
		title = {(a) Network RTT (ms)}
	]
	
	\addplot+ [boxplot prepared={draw position=1,
		median=0, lower quartile=0, upper quartile=0, upper whisker=0, lower whisker=0},
	] coordinates {};
	\addplot+ [boxplot prepared={draw position=2,
		median=9.63, lower quartile=9.35, upper quartile=10.3, upper whisker=11.28, lower whisker=8.5},
	] coordinates {};
	\addplot+ [boxplot prepared={draw position=3,
		median=13.7, lower quartile=13.6, upper quartile=14.6, upper whisker=18.6, lower whisker=12.6},
	] coordinates {};
	\addplot+ [boxplot prepared={draw position=4,
		median=27.7, lower quartile=27.6, upper quartile=28.6, upper whisker=34.716, lower whisker=25.884},
	] coordinates {};

	%
	%
	\nextgroupplot[
		boxplot/draw direction=y,
		ymajorgrids,
		xmajorgrids,
		ymin=0, ymax=10.5,
		ytick = {0,1,...,10},
		title = {(b)   MPC exe. time (ms)},
	]
	\addplot+ [boxplot prepared={
		median=5.1231, lower quartile=5.0991, upper quartile=5.162, upper whisker=9.9721, lower whisker=5.0399},
	] coordinates {};
	\addplot+ [boxplot prepared={draw position=2,
		median=1.0419, lower quartile=0.93794, upper quartile=1.195, upper whisker=1.2591, lower whisker=0.75793},
	] coordinates {};
	\addplot+ [boxplot prepared={draw position=3,
		median=1.0791, lower quartile=0.88882, upper quartile=1.1208, upper whisker=1.467, lower whisker=0.83709},
	] coordinates {};
	\addplot+ [boxplot prepared={draw position=4,
		median=0.55504, lower quartile=0.54812, upper quartile=0.57793, upper whisker=0.67711, lower whisker=0.54502}, ,
	] coordinates {};

	%
	%
	\nextgroupplot[
		boxplot/draw direction=y,
		ymajorgrids,
		xmajorgrids,
		ymax=250,
		ytick={0, 50, ..., 250},
		title = {(c)   Control latency (ms)},
	]
	\addplot+ [boxplot prepared={
		median=11.9493, lower quartile=11.8897, upper quartile=12.7447, upper whisker=24.1734, lower whisker=11.8115},
	] coordinates {};
	\addplot+ [boxplot prepared={draw position=2,
		median=83.4904, lower quartile=69.4618, upper quartile=88.1093, upper whisker=107.9306, lower whisker=55.8602},
	] coordinates {};
	\addplot+ [boxplot prepared={draw position=3,
		median=84.4576, lower quartile=74.5587, upper quartile=95.7839, upper whisker=134.7477, lower whisker=58.1975},
	] coordinates {};
	\addplot+ [boxplot prepared={draw position=4,
		median=135.3097, lower quartile=132.5737, upper quartile=136.9472, upper whisker=225.9486, lower whisker=99.8634},
	] coordinates {};

	%
	%
	\nextgroupplot[
		boxplot/draw direction=y,
		ymajorgrids,
		xmajorgrids,
		ymin=-0.1, ymax=0.7,
		title = {(d)   $u^2$},
	]
	\addplot+ [boxplot prepared={
		median=0.0051735, lower quartile=0.0013428, upper quartile=0.015717, upper whisker=0.20932, lower whisker=2.2382e-06},
	] coordinates {};
	\addplot+ [boxplot prepared={draw position=2,
		median=0.0078458, lower quartile=0.0019422, upper quartile=0.024385, upper whisker=0.29562, lower whisker=3.1914e-06},
	] coordinates {};
	\addplot+ [boxplot prepared={draw position=3,
		median=0.0094932, lower quartile=0.0023403, upper quartile=0.029512, upper whisker=0.42161, lower whisker=4.1418e-06},
	] coordinates {};
	\addplot+ [boxplot prepared={draw position=4,
		median=0.017392, lower quartile=0.0036914, upper quartile=0.050267, upper whisker=0.57808, lower whisker=6.7842e-06},
	] coordinates {};

	\end{groupplot}

\end{tikzpicture}
			\caption{Statistical summary of the \ac{MPC} baseline measurements.}
			\label{fig:mpc_baseline}
		\end{figure*}
		\tikzexternaldisable
	
		Figure \ref{fig:mpc_baseline}a shows the \acp{RTT} from the Raspberry Pi at the plant to the other systems.
		Notably the wireless link realised with \ac{LuMaMi} introduces a latency of $5 ms$ one way, as made evident by the $10 ms$ \ac{RTT} between the plant and the edge node.
		5G is pushing for even faster \ac{RTT} but this is good radio link performance compared to commercially available alternatives.

		\Cref{fig:mpc_baseline}b shows the \ac{MPC} execution time.
		We chose a simple scenario where all nodes can execute the \ac{MPC} with a significant margin. 
		However, clearly, the Raspberry Pi at the plant is many times slower than the other systems.
		The AWS node is faster than edge and ERDC which is to be expected by the specification in \ref{sec:syssetup}.
		We see in this graph that there are large outliers in terms of execution time at the plant.
		Even though we execute in real-time mode we have seen recurrent extended system interruptions to the \ac{MPC} on the Raspberry Pi.
		This could be the cause of these outliers.
		On the \acp{DC} we do not expect real-time properties to apply  outside the virtual machine and can therefore expect some outliers.
		Due to the short execution times, we expect them to be unlikely.
		
		\Cref{fig:mpc_baseline}c shows the aggregate latency from reading the position of the ball to applying the control signal (i.e. adjusting the velocity of the beam).
		This is an important measure because the controller is designed with the assumption that the input to output is instantaneous and hence, that the state of the system has not changed when the control signal is applied.
		We see that the differences in delay are not as pronounced as in \Cref{fig:mpc_baseline}a.
		The execution times in \Cref{fig:mpc_baseline}b and the network latency in  \Cref{fig:mpc_baseline}a are not the only contributors to the control latency.
		This tells us that a significant proportion of the delay in our system is introduced by the software platform or application design and not the network.
		The system dynamics causing the increasing variance in \Cref{fig:mpc_baseline}c is also an interesting topic for further research.
		The effect on our process due to these properties are visualised in \Cref{fig:mpc_baseline}d where we see that the energy of the control signal $u$ increases the further we move from the plant.
		Notice that the AWS node performs well but network delays causes it to exhibit a larger mean and variance in the control signal.
		Such an effect can be part of the heuristics when deciding where to place control in the edge cloud.

	\subsection{System adaptability}\label{ref:adapt}
		We have established that a controller can successfully be implemented on the edge cloud test-bed and studied characteristics in terms of execution times, latencies and jitter.
		Essential to the mutability of the system is its ability to migrate applications and actors to respond to the applications' and the infrastructure's changing objectives.
		During a migration the Calvin cluster performs the necessary modification of the network communication path, recreates the actor at the target node, copies state and handles the transition of token queues.
		Although the actor moves point-to-point, changing the communication paths may involve many nodes in the cluster.
		We now add this perspective by way of relocating the \ac{MPC} amongst the nodes while balancing and repositioning the ball as in Section \ref{ref:character}.

\begin{figure*}[h!]
			\centering 
			\begin{tikzpicture}[scale=0.95,transform shape]
	\pgfplotsset{width=1.025\linewidth,height=5.0cm}
	\pgfplotsset{filter discard warning=false}
	\pgfplotsset{grid style={dashed}}
	
	\begin{groupplot}
	[
		group style={
			group name=my plots,
			group size=1 by 3,
			xlabels at=edge bottom,
			xticklabels at=edge bottom,
			ylabels at=edge left,
			yticklabels at=edge left,
			vertical sep=0.3cm,
			horizontal sep=0.5cm
			},
		ylabel style = {align=center},
		xmajorgrids,
		ymajorgrids,
		xlabel={Time (s)},
		scaled ticks=false,
		xmin=0, xmax=800,
	]

	\nextgroupplot[
	ylabel={(a)\\ $u^2$},
	]
	
	\addplot[] table[x index=0, y expr=\thisrowno{1}*\thisrowno{1}, col sep=comma]{data/migration_ts/downsampled_mpc_action.csv};
	
	\nextgroupplot[
	ylabel={(b) \\ Inputs (cm)},
	legend columns = 2,
	height=3.0cm,
	ytick={0,15,30},
	legend style={at={(1,-1.6)},anchor=south east}
	]
	
	\addplot[red] table[x index=0, y expr=\thisrowno{1}*5.5, col sep=comma]{data/migration_ts/downsampled_pos_trigger.csv};
	\addlegendentry{Position}
	
	\addplot[black, ultra thick] table[x index=0, y expr=\thisrowno{1}*5.5, col sep=comma]{data/migration_ts/downsampled_dis_clocktick.csv};
	\addlegendentry{Set-point}

	\nextgroupplot[
	ylabel={(c) \\ Node},
	height=2.5cm,
	ytick={1,2,3,4},
	yticklabels={\scriptsize{Plant}, \scriptsize{Edge}, \scriptsize{\ac{ERDC}}, \scriptsize{AWS}},
	]
	
	\addplot[ultra thick] table[x index=0, y index=5, col sep=comma]{data/migration_ts/downsampled_mpc_action.csv};

	\end{groupplot}
	
\end{tikzpicture}
			\caption{Time-series of \ac{MPC} being randomly migration between the system's four nodes.}
			\label{fig:mpc_baseline_ts}
		\end{figure*}
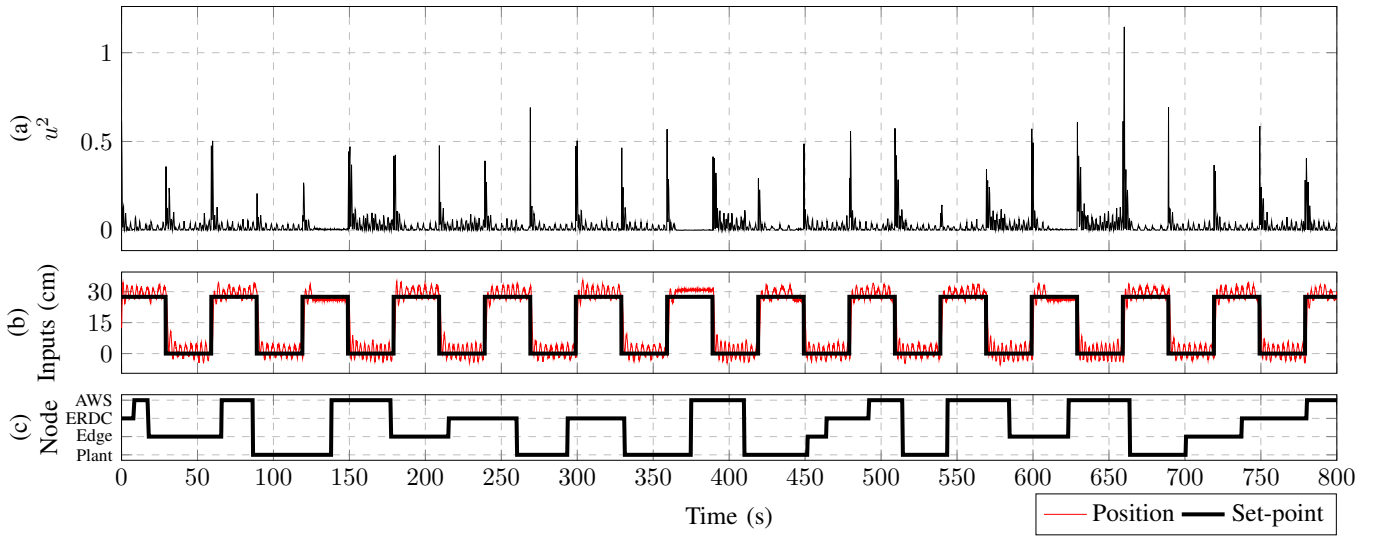

		In \Cref{fig:mpc_baseline_ts} we continuously migrate the \ac{MPC} actor randomly across the four compute nodes, in run-time.
		When doing this, the system must ensure to keep the Kalman filter, the set-point, the previous state\per{There was a comment in the reviews on how much data needs to be transferred and in addition, more info in the real-time requirements. We have not addressed this further. Is that ok?}, and tracing meta data intact.
		Delays, data loss or duplication, and incorrect state transfer negatively impacts the control performance.
		\Cref{fig:mpc_baseline_ts}c shows the placement of the actor in time. 
		\Cref{fig:mpc_baseline_ts}b and \Cref{fig:mpc_baseline_ts}a show the controller inputs and outputs respectively.

		\Cref{fig:mpc_baseline_ts}b shows that the process is stable and is able to operate without interruptions.
		The ball stays on the beam and close to the desired position.
		\Cref{fig:mpc_baseline_ts}a confirms what is presented in \Cref{fig:mpc_baseline}d, i.e., the control signal increases as a function of the distance to the plant.
		Set-point changes are clearly visible has high peaks but the migrations in \Cref{fig:mpc_baseline_ts}c are not evident in \Cref{fig:mpc_baseline_ts}a nor \Cref{fig:mpc_baseline_ts}b. 
		However, the peak in the control signal near the set-point change after 650 seconds is likely caused by a coinciding migration.
		In its current form, the system is not aware of when or to where a migration will occur nor do we attempted to mitigate its potential effects.
	
		\tikzexternalenable
		\tikzexternaldisable
		
	\subsection{Tightened constraints}\label{sec:tight}
		We now demonstrate an example use case where the controller takes advantage of the \paradigm{}.
		Here, relative to the previous example, we have set a constraint on the control signal to the plant.
		Albeit being a synthetic exercise, it is not an unreasonable action since limits in control signal are commonly used to, for instance, reduce actuator wear and to avoid non-linear parts of the operating range.
		To make the associated optimisation problem harder we move the ball just short of the end of the beam.
		In combination with the constraints this will cause a higher load on the \ac{MPC} host node.
		Small disturbances may cause the ball to fall off.
		
\begin{figure}[t!]
			\centering 
			\begin{tikzpicture}[scale=1,transform shape]
	\pgfplotsset{width=0.43\linewidth}
	\pgfplotsset{filter discard warning=false}
	\pgfplotsset{grid style={dashed}}
	
	\begin{groupplot}
	[
		group style={
			group name=my plots,
			group size=3 by 3,
			xlabels at=edge bottom,
			xticklabels at=edge bottom,
			ylabels at=edge left,
			yticklabels at=edge left,
			vertical sep=0.3cm,
			horizontal sep=0.3cm
			},
		ylabel style = {align=center},
		xmajorgrids,
		ymajorgrids,
		xlabel={Time (s)},
		scaled ticks=false,
		xmin=0, xmax=80,
		xtick={5,20,35,50,65,80},
	]

	%
	%
	\nextgroupplot[
	ylabel={Inputs (cm)},
	legend columns = 3,
	height=4.0cm,
	ymax=60, ymin=-10,	
	title = {Plant}
	]
	
	\fill [red, fill opacity=0.25] (axis cs:72.5,-60) rectangle (axis cs:80,60);
	
	\addplot[red] table[x index=0, y expr=\thisrowno{1}*5.5, col sep=comma]{data/constrained/downsampled_pos_trigger_plant.csv};

	\addplot[black, ultra thick] table[x index=0, y expr=\thisrowno{1}*5.5, col sep=comma]{data/constrained/downsampled_dis_clocktick_plant.csv};

	\nextgroupplot[
	legend columns = 3,
	height=4.0cm,
	ymax=60, ymin=-10,
	title = {Edge}
	]
	
	\addplot[red] table[x index=0, y expr=\thisrowno{1}*5.5, col sep=comma]{data/constrained/downsampled_pos_trigger_edge.csv};
	
	\addplot[black, ultra thick] table[x index=0, y expr=\thisrowno{1}*5.5, col sep=comma]{data/constrained/downsampled_dis_clocktick_edge.csv};

	\nextgroupplot[
	legend columns = 4,
	height=4.0cm,
	ymax=60, ymin=-10,
	title = {AWS},
	legend style={at={(1,1.6)},anchor=north east},
	]
	
	\fill [red, fill opacity=0.25] (axis cs:70.5,-60) rectangle (axis cs:80,60);
	
	\addplot[red] table[x index=0, y expr=\thisrowno{1}*5.5, col sep=comma]{data/constrained/downsampled_pos_trigger_aws.csv};
	\addlegendentry{Position}
	
	\addplot[black, ultra thick] table[x index=0, y expr=\thisrowno{1}*5.5, col sep=comma]{data/constrained/downsampled_dis_clocktick_aws.csv};
	\addlegendentry{Set-point}

	\addlegendimage{line width=2mm,color=red, opacity=0.25}
	\addlegendentry{Off beam}
	
	\addlegendimage{mark=o,only marks, blue, thick}
	\addlegendentry{No solution}

%
%
%
	%
	%
	\nextgroupplot[
	ylabel={Exec. (ms)},
	height=4.0cm,
	ymin=0, ymax=100,
	]
	\addplot[] table[x index=0, y expr=\thisrowno{3}*1e3, col sep=comma]{data/constrained/downsampled_mpc_action_plant.csv};
	\draw[blue, thick] (axis cs:72.5,95.5) circle (3pt);
	
	\nextgroupplot[
	height=4.0cm,
	ymin=0, ymax=100,
	]
	\addplot[] table[x index=0, y expr=\thisrowno{3}*1e3, col sep=comma]{data/constrained/downsampled_mpc_action_edge.csv};
	\draw[blue, thick] (axis cs:40,75) circle (3pt);
	
	\nextgroupplot[
	height=4.0cm,
	ymin=0, ymax=100,
	]
	\addplot[] table[x index=0, y expr=\thisrowno{3}*1e3, col sep=comma]{data/constrained/downsampled_mpc_action_aws.csv};
	
	\draw[blue, thick] (axis cs:12.5,40) circle (3pt);
	\draw[blue, thick] (axis cs:15.25,40) circle (3pt);
	\draw[blue, thick] (axis cs:18,40) circle (3pt);
	\draw[blue, thick] (axis cs:20,40) circle (3pt);
	\draw[blue, thick] (axis cs:70,40) circle (3pt);
	\draw[blue, thick] (axis cs:71.5,40) circle (3pt);

%
%
%
	%
	%
	\nextgroupplot[
	ylabel={Latency (ms)},
	height=4.0cm,
	ymin=0, ymax=520,
	]
	\addplot[] table[x index=0, y index=1, col sep=comma]{data/constrained/downsampled_e2e_plant.csv};
	
	\nextgroupplot[
	height=4.0cm,
	ymin=0, ymax=520,
	]
	\addplot[] table[x index=0, y index=1, col sep=comma]{data/constrained/downsampled_e2e_edge.csv};
	
	\nextgroupplot[
	height=4.0cm,
	ymin=0, ymax=520,
	]
	\addplot[] table[x index=0, y index=1, col sep=comma]{data/constrained/downsampled_e2e_aws.csv};

	\end{groupplot}
	
\end{tikzpicture}
			\caption{Time-series of experiments run with tightened constraints on the plant, egde, and \ac{AWS} nodes.}
			\label{fig:upp_pressure}
		\end{figure}
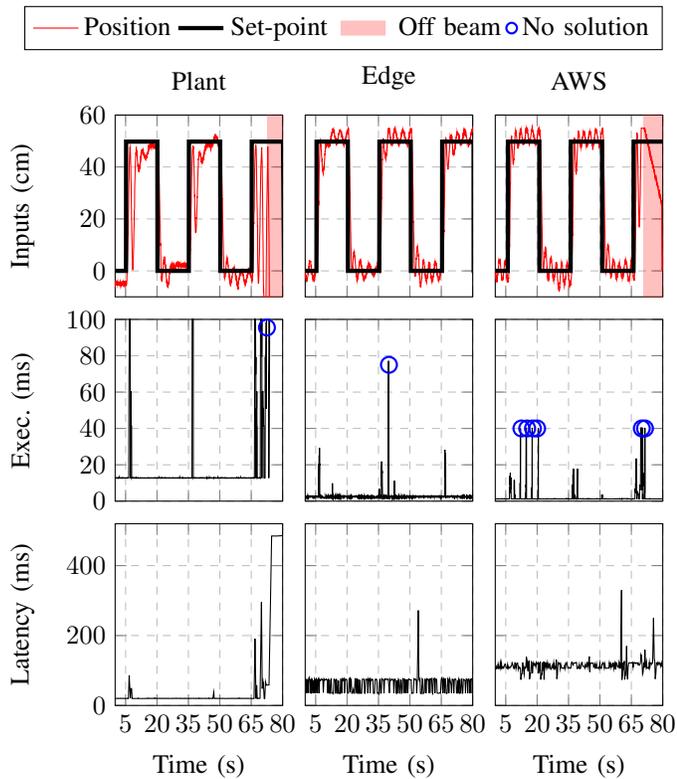

		The experiment shown in \Cref{fig:upp_pressure} applies this configuration. 
		The graphs show time series of the inputs to the controller, the execution time of the \ac{MPC}, and the total latency from input to output of the measured nodes. 
		The blue circles mark occasions when the \ac{MPC} fails to find a solution.
		
		As constraints are tightened it becomes increasingly hard to find a control signal sequence which moves the system from the present state to the target state, while staying within the bounds. 
		This manifests in longer execution times for the optimisation.
		As seen in the execution times in \Cref{fig:upp_pressure}, when the system has settled around a set-point, the optimisation is easy to solve and computationally light.
		In these situations the controller on the plant performs well.
		Latency and jitter of the networked controllers may cause them to deviate more from the set-point.
		\william{Vi borde skriva någonstanns hur vi utvärdear regulatorns prestanda.}
		\per{I linje med efterfrågan om mera information om realtidskrav. Vi diskuterade detta och kom fram till att vi medvetet inte ställer några större krav utan har nöjt oss med att kulan stannar på bomen. Vi har inte heller utvärderat exakt vad som krävs för att den ska stanna kvar då syftet är att undersöka om scenrariot är görbart samt ge en grundläggande bild av (systemets) prestanda.}

		Eventually however, the computational limitations of the plant cause the ball to falls off the beam as a set-point change occurs.
		Note that the plant is not unable to move the ball to the position at the end of the beam but eventually, model errors and noise become too large for it to handle.
		In contrast, the edge node is able to operate without failure.
		Also note that on the \ac{AWS} instance, despite its computational capacity, the controller fails to cope with the resulting latency and system jitter - the ball falls off.
		
		The execution time at the plant when the \ac{MPC} fails to find a feasible solution, is close to an order of magnitude that of the sampling time, represented by a line which extends well beyond the top of the graph.
		This is representative of the computational problems experienced at the plant due to noise during a set-point change.
		In contrast, an equal number of iterations consumes 80 ms on the edge node and only 40 ms on the \ac{AWS}.
		
		With the position closer to the end of the beam and with reduced range in the control output signal, the controller repeatedly experiences non-trivial situations which require additional iterations of the optimisation loop.
		At times, noisy readings make a tough situation even worse and there may seemingly be no solution that keeps the ball on the beam.
		When a new evaluation can be made quickly enough then the state of the system may still be such that the ball can be saved. 
		On the \ac{AWS} node the communication delays increase the frequency of these tough situations and we see repeated problems of finding a solution.
		However, the speed of the \ac{AWS} node allows it to cope with many of these situations since the combined execution and communication delay is much less than the compute time at the plant. 
		Only our edge node is in a position where it is able to handle the full range of the system noise.
		
		\tikzexternalenable
		\tikzexternaldisable
\section{Conclusions}\label{sec:conclusions}
	In this paper we presented an \paradigm{} research test-bed for an IoT and heterogeneous cloud environment.
	We deployed an automatic control application on the test-bed to act on a time-sensitive and mission-critical process. 
	We evaluated the controller's viability, performance, and system characteristics.
	Our evaluation shows that cloud native control loops can viably be deployed on the edge cloud.
	We operate at a sampling rate of 20 Hz but there is potential for this to be pushed further in the near future.
	We also showed that our controller can benefit from the edge cloud and that the system and the placement of the controller can be dynamically reconfigured in run-time without strictly sacrificing stability.

	To further improve the software platform we see that there remain work to be done on actor scheduling, overheads of message passing, and synchronisation of such a tightly connected system in an inherently uncertain cloud environment.
	With improvements, we can increase the requirements on the application and the system to move towards anticipated future applications which require such a system to be user friendly, self-adaptive, resilient, high performing, and deliver low latency and low jitter.

	Importantly, we conclude that we have an observable test-bed which enables us to continuously operate mission-critical applications while performing targeted experiments.
	We have now arrived at a system which allows broad experimental research of the interplay between the application and the underlying platform.
	Continuations to the work targets novel and established techniques within the fields of control theory, distributed systems, and software engineering.
	
	The project source code is available on GitHub \cite{sourcecode}.


\section*{Acknowledgements}
We would like to thank Andreas Johansson and Martin Nilsson at the Department of Electrical and Information Technology at Lund University for their help with designing and constructing the Raspberry Pi shields needed for our experiments.
This work is funded by the Swedish Research Council (VR) under contract number C0590801, the Lund Center for Control of Complex Engineering Systems (LCCC) also funded by VR, and by the Wallenberg AI, Autonomous Systems and Software Program (WASP) funded by the Knut and Alice Wallenberg Foundation.

\bibliographystyle{IEEEtran}
\bibliography{references}

\end{document}